\documentclass[prb,aps,twocolumn,superscriptaddress,showpacs]{revtex4-1}

\usepackage{graphicx}
\usepackage{epstopdf}
\usepackage[colorlinks=true,linkcolor=black, citecolor=blue, urlcolor=blue, 
    unicode=true]{hyperref}

\begin{document}

\title{Depth dependant element analysis of PbMg$_{1/3}$Nb$_{2/3}$O$_{3}$ using muonic X-rays}

\author{K. L. Brown}
\affiliation{School of Physics and Astronomy, University of Edinburgh, Edinburgh EH9 3JZ, UK}

\author{C. P. J. Stockdale}
\affiliation{School of Physics and Astronomy, University of Edinburgh, Edinburgh EH9 3JZ, UK}

\author{H. Luo}
\affiliation{Shanghai Institute of Ceramics, Chinese Academy of Sciences, Shanghai 201800, China}

\author{X. Zhao}
\affiliation{Shanghai Normal University, Shanghai 200234, China}

\author{J. {-}F. Li}
\affiliation{Department of Materials Science and Engineering, Virginia Tech., Blacksburg, VA 24061, U.S.A.}

\author{D. Viehland}
\affiliation{Department of Materials Science and Engineering, Virginia Tech., Blacksburg, VA 24061, U.S.A.}

\author{G. Xu}
\affiliation{National Institute of Standards and Technology, Gaithersburg, MD 90899, U.S.A.}

\author{P. M. Gehring}
\affiliation{National Institute of Standards and Technology, Gaithersburg, MD 90899, U.S.A.}

\author{K. Ishida}
\affiliation{RIKEN Nishina Center, RIKEN, Wako, Saitama, Japan}

\author{A. D. Hillier}
\affiliation{ISIS Pulsed Neutron and Muon Facility, STFC Rutherford Appleton Laboratory, Didcot OX11 0QX, UK}

\author{C. Stock}
\affiliation{School of Physics and Astronomy, University of Edinburgh, Edinburgh EH9 3JZ, UK}

\date{\today}

\begin{abstract}
The relaxor PbMg$_{1/3}$Nb$_{2/3}$O$_{3}$ (PMN) has received attention due to its potential applications as a piezoelectric when doped with PbTiO$_{3}$ (PT). Previous results have found that there are two phases existing in the system, one linked to the near-surface regions of the sample, the other in the bulk. However, the exact origin of these two phases is unclear. In this paper, depth dependant analysis results from negative muon implantation experiments are presented. It is shown that the Pb content is constant throughout all depths probed in the sample, but the Mg and Nb content changes in the near-surface region below 100$\mu$m. At a implantation depth of 60$\mu$m, it is found that there is a 25\% increase in Mg content, with a simultaneous 5\% decrease in Nb content in order to maintain charge neutrality. These results show that the previously observed skin effects in PMN are due to a change in concentration and unit cell.
\end{abstract}

\maketitle


In the conventional theorem of phase transitions, a single temperature-dependant length scale is required to describe the critical fluctuations as the phase transition is approached~\cite{Cowley}. However, a number of systems have been found experimentally which exhibit two length scales at phase transitions, with the second length scale appearing at temperatures just above the critical temperature~\cite{Cowley, Andrews, Shapiro, Hlinka, Wang, Gehring1, Thurston}. The first material for which this was observed was SrTiO$_{3}$, which was found to exhibit an anisotropic dispersion in the critical scattering at the R point~\cite{Andrews}. Similar results have been found in pure Ho and Tb crystals, as well as other compounds~\cite{Gehring1, Thurston, Hirota}. The magnetic analogue to this structural effect has also been observed in spin systems described by the random-field Ising model\cite{Hill1, Hill2}.

Neutron scattering results on these systems have revealed that, of the two phase transitions observed experimentally, one is only present in the bulk, and the other is in the surface of the sample~\cite{Gehring1, Hlinka, Wang, Hirota}. The question remains about whether or not these surface effects are intrinsic properties of the system.


X-ray and neutron scattering studies of relaxors; ferroelectric materials that exhibit large electrostriction; have suggested the presence of an effect similar to the two length scale problem, known as the ``skin effect'', which has a different temperature dependance, appearing far above the critical temperature, and persisting below it~\cite{Xu3}. A number of relaxors are found to have a perovskite structure A(B',B'')O$_{3}$, where there are two different elements sharing the B site, not necessarily with equal proportions of each. Examples of these include PbZn$_{1/3}$Nb$_{2/3}$O$_{3}$ (PZN) and PbMg$_{1/3}$Nb$_{2/3}$O$_{3}$ (PMN), which is the focus of this paper. The key property of a relaxor is their broad frequency dependant dielectric response, which differs from that seen in regular ferroelectrics, as the peak of this dielectric response is not associated with a ferroelectric phase transition~\cite{Colla,Fu}. These materials have been highly researched due to their promising properties as piezoelectric devices when they are doped with PbTiO$_{3}$ (PT), so two competing phases could have a large impact on these properties.


Initially, the phase diagrams of PZN and PMN were studied. PZN systems exhibit a diffuse structural transition between cubic and rhombohedral phases in a temperature range of 325 to 385 K\cite{Iwase, Lebon}. Conversely, PMN does not exhibit a structural phase transition and retains a cubic unit cell down to 5 K, however a strain is observed that is correlated with a phase transition temperature that is predicted from studies of PMN doped with PbTiO$_{3}$\cite{Lebon, Xu3}. Further experimentation found that the Bragg peaks observed in the high temperature cubic phase begin to split at the temperature where this strain manifests, meaning that a cubic and rhombohedral phase are coexisting\cite{deMathan, Gehring2, Stock1}. There is disagreement as to exactly where these two phases are originating from. Some results suggest that the structure is in fact rhombohedral, and the observed effects are as a result of domain size and population gradient within the crystal\cite{Kisi}. On the other hand, several studies suggest a near-surface rhombohedral region which creates the skin effect, around the cubic bulk, with the surface region found to have a depth of approximately 100 $\mu$m\cite{Gehring2, Conlon}. The most profound evidence for this skin effect is the conflict between bulk measurements, such as neutron diffraction, which report a cubic structure, and the rhombohedral structure found from surface-sensitive probes, such as x-ray diffraction\cite{Stock2, Dkhil}. This near-surface region has also been reported in PZN, with a depth of 10 to 50 $\mu$m, where it is suggested that the structural phase transition observed in this system does not necessarily affect the whole bulk\cite{Xu1, Xu2}. Similarly to the two length scale effect, there is not yet any evidence as to the origin of the skin effect, and whether it is intrinsic or not.


In this paper, negative muons are used to provide a depth-dependant compositional analysis of PMN, with the aims of establishing whether there is a difference in composition in the near-surface and bulk regions. This will then establish whether the occurrence of the proposed skin effects is due to sample-specific phenomena, or whether these observations are due to more fundamental properties.  
The use of negative muons as a technique is currently uncommon, despite its effectiveness in characterising samples both in the bulk and near-surface regions. The key advantage of the use of negative muons, as opposed to EXAFS, is that EXAFS is a destructive technique (depending on the surface), making negative muons a preferred method for samples of importance.

\begin{figure}[t]
\includegraphics[width=8.6cm] {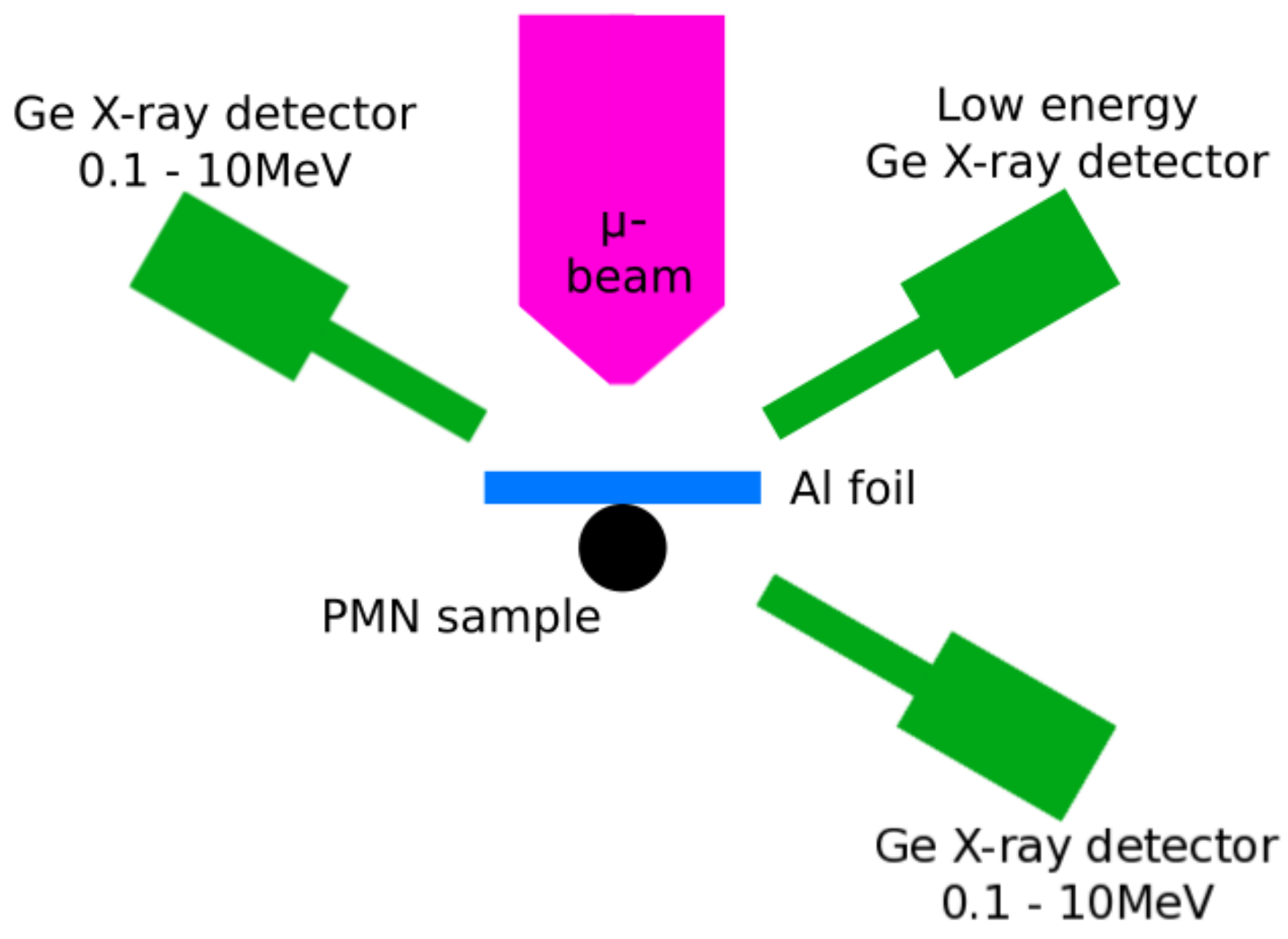}
\caption{\label{ExptSetup} A diagram to show the basic setup of this experiment, showing approximate positions and angles of the equipment used. The sample was centred in the muon beam to achieve maximum bombardment.}
\end{figure}

When a muon is implanted in the sample, it can be captured by one of the atoms in the sample and comes to rest on one of the electronic energy levels. By then measuring the energy of the X-rays emitted when the muon decays, the exact atom and electronic state the muon came to rest upon can be identified~\cite{Engfer, Measday}. This allows for simple calibration and comparison of the data collected. For this experiment, two relevant energy regions were identified: 200 - 500keV, where Mg and Pb peaks are measurable; and 2000 - 3000keV, where Nb and Pb peaks may be measured. These shall henceforth be referred to as the low and high energy regions.


Negative muon experiments were carried out at ISIS using the experimental area CHRONUS on the RIKEN beamline. For these experiments a constant incident momenta of negative muons of 30MeV/c was used, with a 4\% momentum bite (calculated as $\Delta p / p$). A 12mm Pb collimator was used to direct the beam. Three detectors were available for use: low energy and 0.1-10MeV Ge X-ray detectors directed at the front of the sample (next to the $\mu^{-}$ beam), and a second 0.1-10MeV Ge detector directed at the back of the sample. The detector employed in these measurements was the 0.1-10MeV Ge X-ray detector in front of the sample, as the sample was too large for an appropriate intensity of X-rays to reach the detector behind the sample. For a full diagram of the equipment setup see Fig.~\ref{ExptSetup}.

In the results presented here, a large 9.3cm$^{3}$ single crystal of PbMg$_{1/3}$Nb$_{2/3}$O$_{3}$ grown using the modified Bridgemann technique described in Ref.~\onlinecite{Luo} was used. This sample has also been used in previous studies of this system, and as a result has been previously heated and cooled, with the maximum temperature it has been exposed to being 600K (Refs.~\onlinecite{Conlon, Stock1}). A large [100] cut surface was oriented to face the muon beam.

The implantation depth of the muons was controlled using 0.1mm layers of Al foil, in direct contact with the sample, as opposed to varying the energy of the muons as in Ref. \onlinecite{Hillier}. The foil acts to impede the momentum of the incident muons, thus reducing the depth at which they are implanted. By varying the thickness of the foil from 6$\mu$m to no foil, multiple depths in the sample were probed\cite{Data1, Data2}. The implantation depth was calculated using the SRIM and TRIM software to find the stopping profile of Al~\cite{SRIM}, however these implantation depths are not exact as the calculations assume a perfectly monochromatic beam.


\begin{figure}[t]
\includegraphics[width=8.6cm] {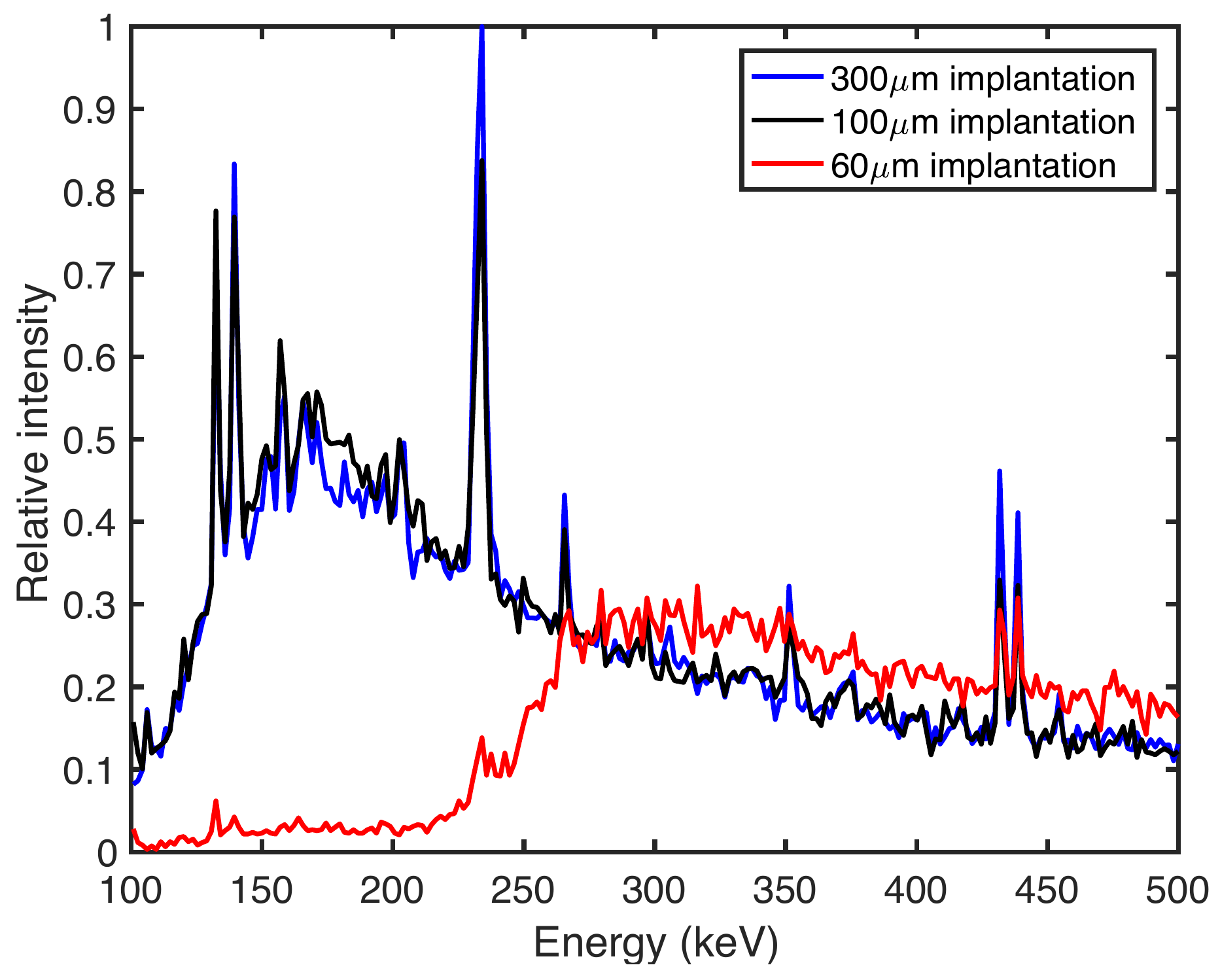}
\caption{\label{RawDat} Graph showing the raw data between 100 and 500keV for the 300$\mu$m, 100$\mu$m and 60$\mu$m data.}
\end{figure}

\begin{figure}[h!]
\includegraphics[width=8.6cm] {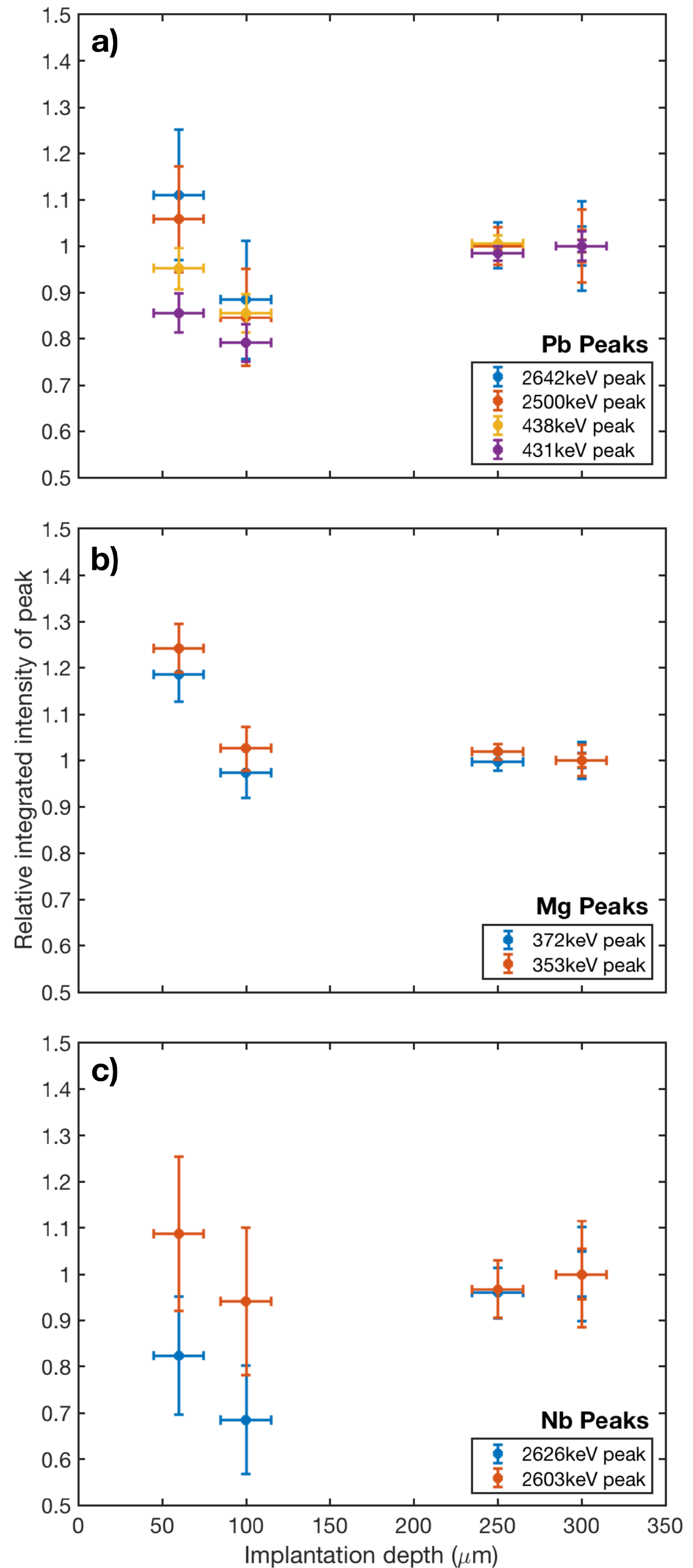}
\caption{\label{Concentrations} The change in relative peak size as the implantation depth is varied for: (a) four Pb peaks; (b) two Mg peaks; (c) two Nb peaks. It should be noted that the errors in implantation depth are systematic.}
\end{figure}

Five different implantation depths were measured, and Fig.~\ref{RawDat} shows the raw data at low energies for three of these measurements, including the largest and smallest penetrations. The peaks examined for the other elements are shown in Table~\ref{PeakTable}, with multiple excitations examined for each. The peak positions have been calibrated to the position of the Pb peaks, which were found to be present in all measurements regardless of depth. We do not believe that there is any low-energy attenuation of the X-rays due to the presence of Pb, because no peak position changes are observed with depth of muon implantation, and all of the measurements are taken very close to the surface so there is only a very small amount of Pb in the path of the X-rays. The measurement closest to the surface shows both the largest background and a cutoff in energy at approximately 250keV. It is for this reason that no results concerning the O concentration is presented, as the peaks for O occur below this energy at 133keV. The results for O at all other depths are found to be constant, so no change in the concentration is observed.

In order to compare the relative concentrations of each element, the integrated intensities of each peak are normalised such that, at the deepest depth probed, which we assume to be the bulk, the relative intensity is 1. For each of the peaks studied the integration range was chosen by hand in order to ensure that there is minimal contribution from background or other peaks. In these experiments no explicit background subtraction was possible, however, estimates of the backgrounds for all depths are found to be comparable, so this is not expected to influence the final results substantially. The resulting plots showing the changes in relative peak intensity for Pb, Mg and Nb are shown in Fig.~\ref{Concentrations}. These changes in intensity are directly linked to the relative abundances of each element. From these results it is then easy to track any changes in concentration with depth.

\begin{table}
\caption{The energies of the peaks examined in this study and their origin.}
\label{PeakTable}
\begin{tabular}{ |c||c|c|c| }
\hline
Element & Excitation & Energy (keV) \\
\hline
Pb & 3d$_{3/2}\rightarrow$ 2p$_{1/2}$ & 2642 \\
 & 3d$_{5/2}\rightarrow$ 2p$_{3/2}$  & 2500 \\
 & 5g$_{7/2}\rightarrow$ 4f$_{5/2}$ & 438 \\
 & 5g$_{9/2}\rightarrow$ 4f$_{7/2}$ & 431 \\
 Mg & 4p $\rightarrow$ 1s & 372 \\
 & 3p $\rightarrow$ 1s & 353 \\
Nb & 2p$_{3/2}\rightarrow$ 1s$_{1/2}$ & 2626 \\
 &  2p$_{1/2}\rightarrow$ 1s$_{1/2}$ & 2603 \\
\hline
\end{tabular}
\end{table}

The changes in concentration for Pb are shown in panel (a) of Fig.~\ref{Concentrations}, and here no clear changes in intensity are seen with depth. This result for Pb is expected to be influenced by the presence of Pb further up in the beam, but we believe that this is not a problem as it is the relative change in peak size being studied here. It is noted that the measurement for an implantation depth of 100$\mu$m appears to be lower than the rest, however this trend appears to be repeated for Nb, so we suggest that this is a result of the total number of muons implanted in the sample being slightly reduced compared to the larger depths due to stopping in the air and aluminium foil. Panel (b) shows the changes in Mg concentration, which shows a clear increase of approximately 25\% at a depth of 60$\mu$m, though the exact depth at which this increase begins is unclear if the 100$\mu$m peak is assumed to be anomalous. The Nb peak height variations in panel (c) are the most difficult to analyse due to the large errors associated with these results. 

When charge neutrality in the system is considered, if there is a 25\% increase in Mg$^{2+}$ ions, there must be an associated 5\% decrease in Nb$^{5+}$ ions. This is within the errors of the measurement carried out here. For the case of a 25\% increase in Mg, the unit cell formula of the system will become PbMg$_{0.42}$Nb$_{0.63}$O$_{3}$. This is no longer an ideal perovskite structure, as there is more than one atom on the B site, which may explain the transformation from the cubic to rhombohedral phase. In order to maintain the ideal perovskite structure, the unit cell formula must be PbMg$_{0.42}$Nb$_{0.58}$O$_{3}$, which is within the errors on the data presented here. This formula gives a net charge of -0.26$e$, meaning that the region is polar. A change in the number of O$^{2-}$ ions could not compensate for this change in Mg content alone, but could if there was a coupled change in Nb, which is a possibility. Similarly, a decrease in Pb$^{2+}$ ions could compensate for this, but as the mass of Pb is significantly larger than that of the elements it can be assumed that this element has least mobility and can be considered constant across the sample.

Previous studies have suggested regions of Nb:Mg = 1:1 concentration to explain anomalous results in diffuse neutron scattering, Raman, and TEM data. These areas are often referred to as polar nano-regions (PNRs), due to the lack of charge balance, and are expected to have rhombohedral symmetry and very small correlation lengths due to their small size\cite{Dkhil, Fu}. A diffraction peak in neutron data was identified as linked to 1:1 concentration regions, and was found to have strong temperature dependance below the phase transition temperature\cite{Gosula}, and other diffraction studies have identified the presence of PNRs from the increase of line tails for temperatures below 600 K\cite{Bonneau}. In Raman studies, a peak was seen at approximately 4 meV and linked to such a 1:1 concentration region, but has not been seen in equivalent neutron studies\cite{Taniguchi, Naberezhnov, Gehring3}. PNRs are expected to also have a large influence upon the ferroelectric soft mode, which has been used to explain some of the unexpected phenomena in the phonon spectra, though it is unclear if this model completely satisfies all of the observations\cite{Stock1, Gehring3, Bosak}. Therefore, it is suggested that a clustering of PNRs near to the surface could be the cause of the observed change in concentration here, as well as the other ``skin'' effects that have been noted in other studies. 

It should also be noted that chemically ordered regions (CNRs) have also been observed in relaxor systems through TEM measurements, which may also be linked to changes in concentration to maintain charge neutrality, but they have been identified to require excess Nb\cite{Fu}. This suggests that, assuming the sample has overall 2:1 Nb:Mg concentration, then CNRs must form in order to compensate for the existence of PNRs. TEM studies have also shown that PNRs inhibit domain growth due to their negative charge, which raises questions as to at what point these regions appeared in the sample\cite{Chen}. No previous studies have reported a gradient in O concentration, so the assumption made here that this remains constant is reflected in the literature.

The results presented here are unable to give evidence for the mechanism behind this change in concentration with depth. However, due to the long length scale for the concentration variation, it is suggested that the cause of the concentration gradient must be driven by diffusion. As these measurements were carried out on a cut and polished surface, the statement that this change in concentration is an artefact of the sample growth cannot be made. Rather, it is possible that the cutting and heating of the sample to create this surface may have influenced this diffusion. The fact that this concentration change is observed suggests that the composition may be close to being unstable, in which case this separation of phases may serve to stabilize the bulk phase. Evidence from other studies have found that surface effects can be much shorter range, with a depth of 10-50$\mu$m reported in PbZn$_{1/3}$Nb$_{2/3}$O$_{3}$~\cite{Xu2}.


To conclude, the results collected here using negative muon techniques show that there is a clear increase in the concentration of Mg in PbMg$_{1/3}$Nb$_{2/3}$O$_{3}$ (PMN) less than 100$\mu$m below the surface, which we also link to a reduction in the Nb concentration. Three possible structures are suggested for the surface region: the rhombohedral PbMg$_{0.42}$Nb$_{0.63}$O$_{3}$; the polar PbMg$_{0.42}$Nb$_{0.58}$O$_{3}$; or a combination of PbMg$_{1/3}$Nb$_{2/3}$O$_{3}$ and polar nano-regions with 1:1 Nb:Mg concentration. The presence of polar nano-regions has been previously reported in other studies, and we thus conclude that our results show a clustering of 1:1 Mg:Nb regions near the surface of the crystal. This suggests that the skin effects observed in this compound are linked to the changes in unit cell and the associated strains.

\bigskip
Support is gratefully acknowledged from the EPSRC, the STFC, the Royal Society, and the Carnegie Trust for the Universities of Scotland.


%

\end{document}